\documentclass[twocolumn,showpacs,prl,aps]{revtex4}
\usepackage{graphicx}
\usepackage{natbib}

\newcommand{\be}{\begin{equation}}
\newcommand{\ee}{\end{equation}}
\newcommand{\beq}{\begin{eqnarray}}
\newcommand{\eeq}{\end{eqnarray}}

\newcommand{\W}{\Omega}
\newcommand{\g}{\gamma}

\newcommand{\bnn}{\begin{eqnarray*}}
\newcommand{\enn}{\end{eqnarray*}}

\newlength{\textwidthm}
\begin{document}

\title{
	Enhancement of field generation via maximal atomic
	coherence prepared by fast adiabatic passage in Rb vapor
}

\author{
	V.~A.~Sautenkov$^{1,3}$,
	C.~Y.~Ye$^{1}$,
	Y.~V.~Rostovtsev$^{1}$, 
	G.~R.~Welch${^1}$, and
	M.~O.~Scully$^{1,2}$
}

\affiliation {
	${^1}$Institute for Quantum Studies and Department of Physics,
	Texas A\&M University, College Station, TX 77843\\
	$^{2}$Department of Chemistry, Princeton University,
	Princeton, NJ 08544\\
	${^3}$Lebedev Institute of Physics, Moscow 117924, Russia
%
}

\date{\today}

\begin{abstract}
\vskip12 pt

We have experimentally demonstrated the enhancement of
coherent Raman scattering in Rb atomic vapor by exciting
atomic coherence with fractional stimulated Raman adiabatic
passage.  Experimental results are in good agreement with
numerical simulations.  The results support the  possibility
of increasing the sensitivity of CARS by preparing atomic or
molecular coherence using short pulses.

\end{abstract}

\maketitle


	Although four-wave mixing processes have been studied
for at least twenty-five years~\cite{boyd,chen}, it has only
recently been shown that quantum coherence can dramatically
increase the nonlinear response in atomic, molecular, or
solid state media without overwhelming 
absorption~\cite{harris,harris2,xio,hemmer,matsko}.
Currently the subject of 
quantum coherence is the
focus of broad research including 
quantum computing and quantum
state storage~\cite{hau2,lukin1,zibrov1}, manipulation of single
quanta~\cite{harris-photon,harris1,lukin2}; 
in a coherently prepared medium with maximal coherence there occur
effective frequency conversion~\cite{jain95prl,hakuta},
sub-femtosecond pulse generation~\cite{sokol},
and enhanced
CARS spectroscopy~\cite{fast-cars}.

\begin{figure}
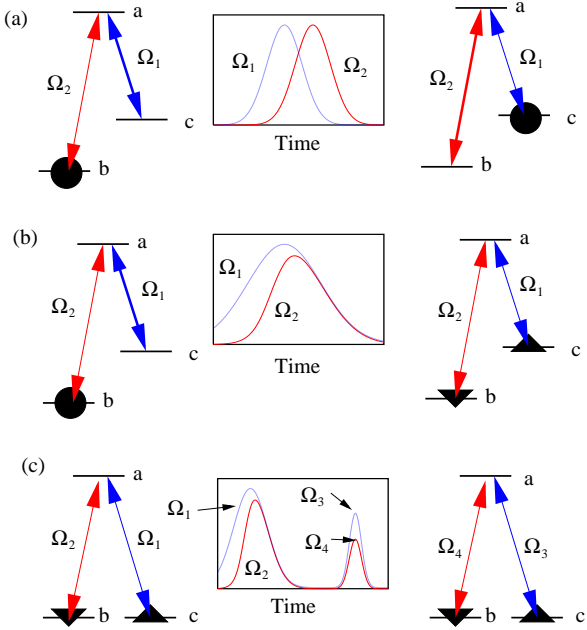
 
\center{
}
\caption{\label{levs-stirap}
	(a) STIRAP. ``Counter-intuitive'' pulse sequence
	transfers population from level $b$ to $c$ with 100\%
	efficiency.
	(b) Fractional STIRAP. A pair of pulses with the same
	back edge distributes populations equally and excites
	maximal coherence between levels $b$ and $c$. The
	coherence between levels is depicted by triangles.
	(c) Applying $\W_1$ and $\W_2$ to a three-level
	atom, one can excite maximal Zeeman coherence via
	fractional STIRAP.  Field $\W_4$ is generated via
	coherent scattering of probe field $\W_3$.
}
\end{figure}

	Maximal coherence in atoms and molecules can be created
with a pair of short coupling pulses which adiabatically
interact with an atomic or molecular system as depicted
in Fig.~\ref{levs-stirap}.  By choosing the proper time
dependence of the pulses shown in Fig.~\ref{levs-stirap}a,
100\% population transfer between lower levels can be achieved.
This is referred to as Stimulated Raman Adiabatic Passage,
or STIRAP~\cite{bergmann,netz}.  However, using the time
dependence shown in Fig.~\ref{levs-stirap}b, referred to as
fractional STIRAP (to distinguish the technique from STIRAP)
we can create maximal coherence between the lower levels.

	In this Letter, we report an experimental implementation
of fractional STIRAP in a Rb vapor cell near room temperature
where lower level coherence is created between Zeeman sublevels.
We demonstrate coherent Raman scattering in the Rb vapor
using short laser pulses (shorter than the lifetime of the
excited state.)  We observe enhancement of the coherent Raman
scattering under the condition of maximal coherence between
Zeeman sublevels prepared by fractional STIRAP.  The observed
intensity of the signal pulse depends quadratically on the
density of Rb atoms.  We theoretically predict this behavior,
and show that the experiments are in agreement with numerical
simulations.

	The essence of the technique is the following.
Two coupling pulses (with duration less than the life-time
of the excited state) with Rabi frequencies $\W_{1}$ and
$\W_{2}$ resonant with transitions $a-c$ and $a-b$ (see
Fig.~\ref{levs-stirap}) create coherence between levels $b$
and $c$.  After some time, (less than the life-time of the
lower level coherence) the probe pulse $\W_3$ arrives and
scatters from the atomic coherence $\rho_{bc}$ leading to
efficient generation of a signal field $\W_4$. The important
points are that the frequency of the signal field is shifted
exactly to the transition frequency between levels $b$ and $c$,
and the intensity of the signal field depends on the magnitude
of the atomic coherence $\rho_{bc}$.

\begin{figure}
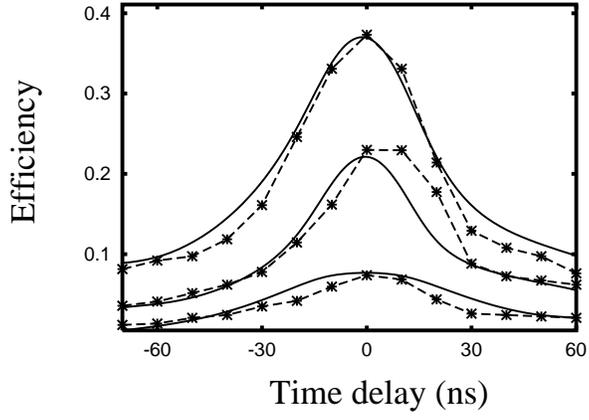
 
\center{
}
\caption{\label{efficiency-fig}
	Efficiency of generation of field $\W_4$ versus mutual
	delay between coupling pulses for three different
	atomic densities.  Zero delay was selected for the
	condition where the tails of the pulses coincide. 
	Experimental results (dotted curve). 
	Results of numerical simulations (solid curve).
}
\end{figure}

	The main result of this Letter is shown in
Fig.~\ref{efficiency-fig}.  By changing the time delay, one
can see that there is a maximum of efficiency of generation
of the signal field corresponding to the configuration of
the coupling pulses that creates maximum coherence.  It is
this coherence that distinguishes the present technique from
ordinary coherent anti-Stokes Raman spectroscopy (CARS) where
the coherence level is low.

	An important feature of our experiments that
distinguishes them from prior experiments such as those on
photon storage~\cite{hau2,lukin1,zibrov1} is that the duration
of the pulses are short in comparison with the relaxation
rates of the associated optical transitions.  Furthermore,
in those experiments the magnitude of $\rho_{bc}$ coherence
was relatively low, and the duration of the pulses was long
compared to the optical lifetime and comparable to the spin
coherence lifetime. Note also that, because
there is no inhomogeneous broadening of spin transition,
our approach is different from the echo technique where the additional 
$\pi-$pulses to control spin coherence are applied~\cite{mossberg}. 

Also our work is differnet from the work done by Harris group at Stanford
where very efficient technique 
employing maximal atomic coherence was developed.  
Their duration of pulse
is longer than the optical relaxation time (duration of the laser pulses 
is $15$ ns, while $T_1 = 5.2$ ns for level $6p7s\; ^3P_1$ of $^{208}Pb$). 
Thus, the laser field is practically quasi-stationary, and 
population of the dark state occurs via optical pumping rather 
than via rapid adiabatic passage.
While in our experiments, we use a fractional STIRAP
to create coherence between hyperfine levels,
also the duration of pulses is shorter than 
population relaxation of the excited state.
Using the regime of strong saturation regime gives 
one a very robust effect of coherence excitation between 
atomic levels. 
%

\begin{figure}
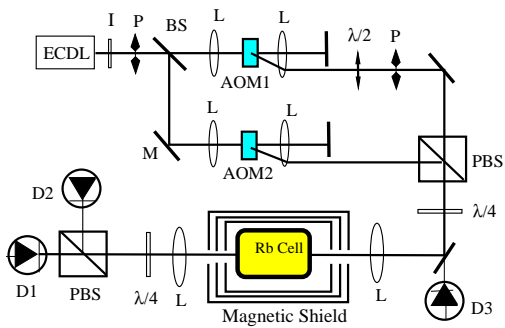
 
\center{
}
\caption{\label{expt-scheme}
	Experimental setup.
}
\end{figure}

 
	A schematic diagram of the experimental setup
is shown in Fig.~\ref{expt-scheme}.  Radiation from an
external-cavity diode laser is tuned to the $D_1$ transition
of $^{87}\mathrm{Rb}$ $5S_{1/2}$ $(F=2)$ to $5P_{1/2}$
$(F'=1)$.  The beam is split by a beam-splitter and passes
through acousto-optic modulators (AOM) driven by pulses with
adjustable duration and delay (the rise time 
of the AOM is near $9$~ns), and the same frequency shift (200~MHz).
The optical path lengths for both beams are the
same.  The temporal behavior of the pulses is detected by fast
photodiode $D_3$. 
	
	One beam serves as the first coupling pulse and
the probe pulse.  The time duration of the first coupling
pulse is $150$~ns, and that of the second pulse is $20$~ns.  
The other beam, also of $20~\mathrm{ns}$ duration,
is used as the second coupling pulse.  Front and tail slopes
of the pulses are limited by the AOMs and are synchronized
electronically.  The polarization of the first beam (which
includes the first coupling pulse and the probe pulse) is
rotated by $90^\circ$ relative to the polarization of the
other pulse.  These orthogonally polarized laser beams are
combined by a polarizing beam-splitter.

	The polarization of the pulses is modified by a
$\lambda/4$ wave-plate, which results in opposite circular
polarization of the two pulses.  The combined laser beams are
focused by a lens (focus length 30 cm) into a cell of length
$2.5~\mathrm{cm}$ containing saturated Rb vapor with atomic
density $N = 1\times 10^{11}~\mathrm{cm}^{-3}$.  The cell is
installed in a three-layer magnetic shield.  The atomic density
is estimated from the temperature of the cell and corrected by
absorption measurements~\cite{Sautenkov}.  After the cell, the
transmitted laser beams with opposite circular polarizations
are separated by a second $\lambda/4$ wave-plate and another
polarizing beam splitter.  The transmitted first coupling
pulse and probe pulse are detected by fast photodiode $D_1$.
The second coupling pulse and the generated signal pulse are
detected by fast photodiode $D_2$.  All fast photodiodes have
identical characteristics with $1$~ns resolution.

\begin{figure}
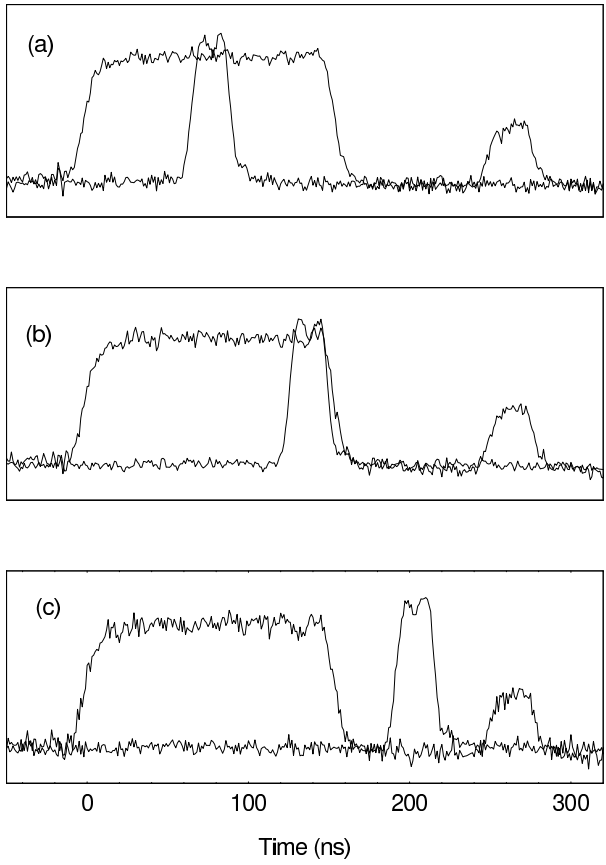
 
\center{
}
\caption{\label{data-pulses}
	Experimental pulse shapes recorded before propagation
	of the optical fields through the Rb cell.  The noise
	level in these data is due to the electronics.
}
\end{figure}

	The coupling pulses create ground state coherence in
the atomic vapor.  Then after some time delay, 
the pulse of probe field scatters on the atomic coherence 
to generate of a new field (signal).
Three different temporal combinations
of the coupling pulses and the probe pulse are shown in
Fig.~\ref{data-pulses}.

	The energy of the first coupling pulse is $375$ pJ, 
its duration is 150~ns that 
was selected so that several optical pumping cycles can occur
before the second coupling pulse is applied. The time delay
between the end of the first coupling pulse and the probe pulse
is 100~ns, which is considerably less than the ground state
coherence decay time of about $10^3$~ns, and more than three
times longer than the excited state lifetime.  
The energy of the second coupling pulse is $50$ pJ,
ans its duration of $20$~ns that is short enough to perform
STIRAP for the time period less than the life time of excited
state ($27$ ns). 
Effective area of the laser beams is about $2\; 10^{-3}$ cm$^2$; 
corresponding 
Rabi frequency for every beam is
$\Omega = 10 \gamma$ (where $\gamma = 1/\tau$ and $\tau = 27$ ns),
so we fulfilled the condition
$\tau\sqrt{\Omega_1^2 + \Omega_2^2} \ge 10$ 
(see in Ref.\cite{bergmann} for details).

%



\begin{figure}
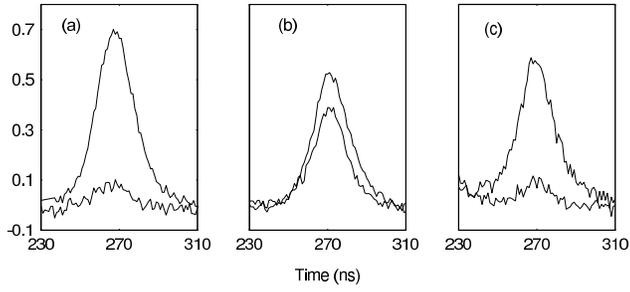
 
\center{
}
\caption{\label{data-generation}
	Probe and generated signal fields after propagation
	through atomic vapor.  In all figures, the probe field
	is stronger than the signal field.  Amplitudes are
	normalized to amplitude of the probe pulse before
	entering the cell.  Fig.~\ref{data-generation}b
	demonstrates the maximal amplitude of the generated
	signal field obtained by the combination of preparation
	pulses shown in Fig.~\ref{data-pulses}b.
}
\end{figure}

	Figure~\ref{data-generation} shows the probe and
signal fields after their propagation through the cell.  The
combination of coupling pulses shown in Fig.~\ref{data-pulses}a
and \ref{data-pulses}c is not optimal for a large ground-state 
coherence to remain after the pulses are gone.  
As a result, the corresponding signal fields in
Fig.~\ref{data-generation}a and \ref{data-generation}c have
small amplitudes.  The optimal signal (with maximum amplitude)
shown in Fig.~\ref{data-generation}b is obtained by adjusting
the time delay of the second coupling pulse relative to the
first coupling pulse in such a way that the condition for
fractional STIRAP is fulfilled.  In this case, fractional
STIRAP induces ground state coherence which is expected to be
close to the maximal coherence.  The amplitude of the generated
field is 0.37 of the amplitude of the initial probe pulse.

	Figure~\ref{efficiency-fig} shows the normalized
amplitude of the generated new field (signal) as a function
of delay time between coupling pulses for three different
densities $N = (1; 0.8; 0.4)\times 10^{11}~\mathrm{cm}^{-3}$.
Zero time delay corresponds the situation where the tails of the
first and second coupling fields switched off simultaneously.
This is the condition for obtaining effective fractional STIRAP
and maximal coherence.  One can see that the efficiency is
proportional to the square of the atomic density.


	To gain physical insight into this process, we have
performed numerical simulations for the propagation of the
laser pulses in the medium.  Instead of the real physical energy
level manifolds ($5S_{1/2}(F=2) \leftrightarrow 5P_{1/2}(F'=1)$
transition) we base our calculations on an idealized three level
system as depicted in Fig.~\ref{levs-stirap}a.  The experiments
are done near room temperature, so we assume that the ground
states are equally populated before the coupling pulses are
applied.

	Consider a three-level system selectively coupled
by three laser pulses.  The first two are coupling pulses,
and the third is a probe pulse.  The two coupling fields,
$\varepsilon_1$ and $\varepsilon_2$, with Rabi frequency
$\Omega_{1}$ and $\Omega_{2}$, respectively, excite transitions
$a \rightarrow c$ and $a \rightarrow b$.  This generates
coherence between levels $b$ and $c$.  Somewhat later, the
probe pulse $\varepsilon_3$, with Rabi frequency $\Omega_{3}$,
interacts with the medium.  Because the medium already has
coherence $\rho_{bc}$, coherent Raman scattering of the probe
pulse leads to the generation of a signal pulse $\varepsilon_4$
with Rabi frequency $\Omega_{4}$.

	The interaction Hamiltonian in the rotating wave
approximation for this system is
\be
H=-\hbar(\sum_{ij}\Omega_{ij}|i\rangle\langle
j|+h.c.)-\hbar\sum_{j}\Delta_j |j\rangle\langle j|,
\ee
where $\Omega_{ij}=\wp_{ij}\varepsilon_{ij}/\hbar$ is the
Rabi frequency of the respective fields, $\wp_{ij}$ is the
electrical dipole matrix element between states $i$
and $j$, $\varepsilon_{ij}$ is the amplitude of the
respective laser field, and $\Delta_{j}$ is the laser detuning
from the atomic resonance.  The time-dependent density matrix
equations of motion are
\be 
\frac{\partial{\rho}}{\partial{\tau}}
=-\frac{i}{\hbar}[H, \rho]-\frac{1}{2}(\Gamma\rho+\rho\Gamma),
\ee
where $\Gamma$ is the decay operator, and $\rho$ is the
atomic density matrix.  Solving these equations gives the time
evolution of the density matrix.  To form a self-consistent
system of equations, one should add an equation for field
propagation which is given by
\be
\frac{\partial{\varepsilon_{ij}}}{\partial{\xi}}=-i\eta\rho_{ij}(\xi,\tau),
\ee
where $\varepsilon_{ij}$ is the Rabi frequency of the field that
is coupled to transition $i \leftrightarrow j$, $\rho_{ij}$
is the coherence between levels $i$ and $j$, $\eta=\nu
N \wp_{ij}/(2\epsilon_0 c)$ is a coupling constant, $\nu$
is the frequency of the field, $N$ is the density of medium,
$\epsilon_0$ is the permitivity of the vacuum, and $c$ is the
speed of light in vacuum.  We use coordinates $\xi$ and $\tau$
which are related to the laboratory coordinates by $\xi=z$
and $\tau=t-z/c$.


	We performed numerical simulations of the above
theory using the same values of the parameters as in the
experiments.  The results of these simulations are shown in
Fig.~\ref{efficiency-fig}(solid curve).  Experimental and theoretical
curves show similar behavior except in the wings, and small
differences are seen in the maxima.  Residual signal at large
delay time can be associated with the long tails of pulses and
residual CW background of the optical fields.

	We have also studied the power dependence of the
generation of the signal field.  The total power of the pulses
was reduced by variable attenuator installed before the first
beam-splitter.  The conversion efficiency varies only slightly
for large changes in power (from our maximum level to half)
and then decreases very rapidly as power is further reduced.
These measurements confirms that we have enough power for
STIRAP.  The good agreement between our experimental and
theoretical results confirms that we have indeed obtained
maximal coherence via fractional STIRAP and observed enhanced
time delayed stimulated Raman scattering with high efficiency.
Fractional STIRAP is a robust technique to obtain maximum Raman
scattering.  We would like to note that we have also studied
delayed Raman scattering at higher Rb densities, but we cannot
obtain efficiency more than $0.4$ due to the strong absorption
and optical pumping to the other hyperfine level ($F=1$).


	The results of our experiment support a femtosecond
adaptive spectroscopic technique for coherent anti-Stokes
Raman spectroscopy (FAST CARS) that has been recently
suggested~\cite{fast-cars}.  The technique takes advantage of
maximum coherence induced between lower atomic or molecular
vibrational levels.  Several methods can be used for effective
preparation of atomic coherence~\cite{fast-cars}, such as
coherent population trapping~\cite{CPT}, chirped pulses,
and the fractional STIRAP~\cite{f-stirap} which we use in this
experiment.  These results constitute the the first experimental
evidence that the FAST CARS technique can be realized.

	Note that FAST CARS is a universal technique which can
be applied to either atoms or molecules in various environments
(gas, liquid, and solids) and can be used in a practical
time scale.  Another important feature of FAST CARS is that
it is capable of determining vibrational frequencies in one
laser pulse. That is significantly different from ordinary
CARS where the frequency of signal field is determined by the
four-photon resonance, and tuning the coupling frequency is
essential to determine the splitting of the levels.  In FAST
CARS, the atomic coherence created by the first pair of pulses
is oscillating at the frequency difference of the lower levels,
and, after some time delay, an anti-Stokes signal appears as
scattering from the probe pulse.  Thus, the frequency shift
between the probe and signal fields is exactly the splitting
between the $b$ and $c$ levels.


	In summary, we have implemented fractional STIRAP in
a Rb vapor cell to generate maximum coherence between Zeeman
sublevels.  We have theoretically predicted and experimentally
demonstrated coherent Raman scattering using short laser pulses
(shorter than the relaxation of lifetime of excited state).
We observe an enhancement of coherent Raman scattering under the
condition of maximal coherence prepared by fractional STIRAP
in agreement with the numerical simulations.  The observed
intensity of the signal pulse depends quadratically on the
density of Rb atoms.  These results support the idea of a FAST
CARS technique utilized in a femtosecond time scale in order
to improve sensitivity of CARS measurements and can be applied
to various atomic, molecular, and biological systems.


	We thank K.K.~Lehman, R.~Lucht, A.B.~Matsko, 
S.E.~Harris, P.R.~Hemmer, W.S.~Warren for useful discussions 
and gratefully acknowledge the support from 
the Office of Naval Research, 
the Air Force Research Laboratory (Rome, NY), Defense Advanced
Research Projects Agency-QuIST, Texas A$\&$M University
Telecommunication and Information Task Force (TITF) Initiative,
and the Robert A.\ Welch Foundation.


\newpage

\section*{Figure Caption}

%





%





%





%


\end{document}